\documentclass[floatfix,twocolumn,prl,longbibliography]{revtex4-2}
\usepackage{graphicx}
\usepackage{dcolumn}
\usepackage{bm}
\usepackage{amsmath}
\usepackage{times}
\usepackage{color}
\usepackage[breaklinks=true,colorlinks,citecolor=blue,linkcolor=blue,urlcolor=blue]{hyperref}

\makeatletter

\newcommand{\Rmnum}[1]{\expandafter\@slowromancap\romannumeral #1@}
\makeatother

\begin{document}

\title{Anisotropic magnetic-field modulation of hysteretic sheet resistance in few-layer ferroelectric WTe$_2$}

\author{Arpan Manna}
\author{Soumik Mukhopadhyay}%
 \email{soumikm@iitk.ac.in}
\affiliation{Department of Physics, Indian Institute of Technology Kanpur, Kanpur 208016, India}

\begin{abstract}

Ferroelectricity in few-layer WTe$_2$, a polar semimetal, originates from interlayer sliding and can be probed through gate-dependent resistance hysteresis. Yet the role of magnetic fields in this transport signature remains largely unexplored. Here, we examine magnetic-field-induced changes in the hysteresis in exfoliated few-layer WTe$_2$. Applying an out-of-plane magnetic field produces a pronounced hysteresis enhancement, while an in-plane field yields a negligible response. Analysis of the individual hysteresis branches reveals distinct magnetotransport responses of the two ferroelectric states, which increase their resistance separation under magnetic field. The enhancement exhibits a robust power-law field dependence and a temperature evolution closely resembling that of magnetoresistance. Our results reveal a close association between anisotropic magnetotransport and ferroelectric hysteresis, underscoring the interplay between magnetic response and ferroic order in a two-dimensional polar semimetal.
\end{abstract}

\maketitle


\section{Introduction}
The ability to manipulate electronic states using multiple external stimuli is of considerable interest for developing multifunctional electronic devices~\cite{Liu_ACSNano2024Integration, Zhang_ADFM2024Emerging}. Materials exhibiting ferroic order alongside electronic transport offer opportunities to combine electrically controllable states with transport-based readout~\cite{Spaldin_NatMater2019Advances, Multiferroics_nm2019march,Scott_Science2007Applications,Jiang_Nano2025Two,Zhang_NatRevMater2023Ferroelectric,Jin_ACSNano2022Ferroelectrics}. In particular, ferroelectric metals provide gate-controlled resistive states that can be distinguished through their transport response~\cite{Fei_Nature2018Ferroelectric, Jia_PRL2025Coexistence,Lu_PRL2019Ferroelectricity,Anderson_PRL1965Symmetry,Ji_APR2026Advances,Zhou_JJAP2020Review}. Exploring how these states respond to additional external stimuli is therefore important for physical understanding of the interplay between ferroelectricity (FE) and electronic transport, and for expanding their potential functionalities~\cite{Wang_NatCommun2025Nonvolatile, Sharma_SciAdv2019Room, delaBarrera_NatCommun2021Direct}.
In this context, the polar semimetal WTe$_2$ has drawn considerable interest, driven by the unique interplay between its broken inversion symmetry, strong spin--orbit coupling, and semimetallic electronic structure~\cite{delaBarrera_NatCommun2021Direct,Chiang_NanoLett2025Spatially,Chen_NatCommun2026Giant,Peng_NatCommun2017Observation,Soluyanov_nature15}. In contrast to conventional ferroelectrics, few-layer WTe$_2$ exhibits robust sliding FE switching despite its semimetallic nature, making it one of the few examples of a ferroelectric van der Waals semimetal~\cite{Lai_NatNano2021Third}. The interplay between FE switching and its associated charge redistributed electronic configurations has driven widespread research, motivated by both fundamental physics and potential applications~\cite{delaBarrera_NatCommun2021Direct, Fei_Nature2018Ferroelectric,Sharma_SciAdv2019Room, Wang_NatCommun2025Nonvolatile, Xue_APR2021Emerging}.

Alongside its sliding ferroelectricity, WTe$_2$ hosts a variety of transport phenomena arising from its strongly anisotropic topological electronic structure, charge compensated semimetallicity, exceptionally large magnetoresistance~\cite{ali2014large, Y.Wang, thoutam2015temperature, zhu2015quantum, rhodes2015role, wang2016breakdown, fatemi2017magnetoresistance, Meng_JPCM2020large, Na_Nanoscale2016tuning, Pan_FrontPhys2017carrier}. Its strongly tilted type-II Weyl nodes, layered crystal structure, and quasi-two-dimensional Fermi surface contribute to a pronounced orientation dependence of its magnetotransport response~\cite{Pletikosic_PRL2014_Electronic,Linnartz}. 
Given the close association between FE switching in WTe$_2$ by interlayer sliding and charge redistribution, the anisotropic magnetotransport provides a potential route to exploring the magnetic-field controllability of its transport-detected FE response~\cite{delaBarrera_NatCommun2021Direct,Fei_Nature2018Ferroelectric, Yang_JPCL2018Origin, Sharma_SciAdv2019Room}.

Recent studies have primarily focused on nonlinear transport, Berry-curvature-driven responses, and electrically switchable nonreciprocal ferroelectricity in WTe$_2$~\cite{Ma2019,Kang_nm19,Tewari_prb21,Tiwari2021, Sodemann_PRL2015Quantum,Yasuda_Science2021Stacking}. However, the direct influence of magnetic field remains comparatively unexplored. In particular, an outstanding question is whether the different electronic environments associated with the two ferroelectric states lead to distinct magnetic-field-dependent transport responses.~\cite{Ye_PRB2025Current}.

In this work, we investigate the magnetic-field dependence of transport-detected ferroelectric hysteresis in few-layer WTe$_2$ devices. Combining magnetic-field-dependent, branch-resolved, and temperature-dependent transport measurements, we demonstrate that the two hysteresis branches exhibit distinct magnetotransport responses, resulting in an enhanced resistance separation under out-of-plane magnetic fields. This enhancement is strongly anisotropic and follows power-law magnetic field dependence. Our findings establish a close association between anisotropic magnetotransport and transport-detected FE hysteresis in few-layer WTe$_2$.

\section{Experimental methods}
\begin{figure*}
    \centering
    \includegraphics[width=\linewidth]{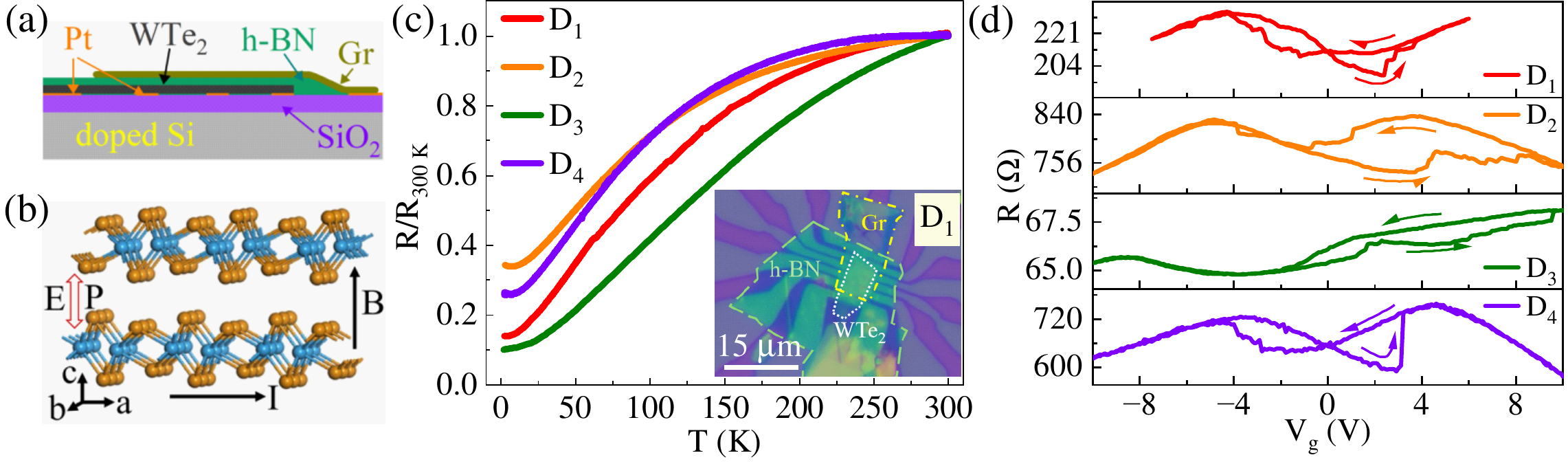}
    \caption{Device structure and transport characterization of top-gated ferroelectric WTe$_2$ heterostructures.
(a) Cross-sectional schematic of the top-gated few-layer WTe$_2$ device architecture and (b) Measurement geometry of the device.
(c) Temperature dependent sheet resistance normalized at room temperature ($R/R(300 K)$), showing metallic behavior for all four devices D$_{1-4}$. Inset: optical microscope image of the fabricated device D$_1$ (The WTe$_2$ channel, metallic contacts, and top-gate region are marked).
(d) Representative gate-voltage-dependent in-plane sheet resistance showing the characteristic FE hysteresis curves obtained during consecutive forward and backward top-gate voltage sweeps and at 2~K for all four devices.}
    \label{MFE_Fig1}
\end{figure*}

Bulk single-crystalline WTe$_2$ crystals were synthesized using the conventional chemical vapor transport (CVT) method~\cite{Linnartz}. Details of the crystal growth and characterization are provided in our previous work~\cite{Manna_MTQuantum2026_Disorder}. Thin flakes of WTe$_2$, hexagonal boron nitride (h-BN), and few-layer graphene/graphite (Gr), serving as the channel, gate dielectric, and gate electrode, respectively, were mechanically exfoliated from bulk crystals onto polydimethylsiloxane (PDMS) stamps using the conventional Scotch-tape method inside an inert glovebox. WTe$_2$ flakes were identified on the PDMS stamps based on their optical contrast, and their thicknesses were tentatively estimated to be below 10~nm~\cite{Na_Nanoscale2016tuning, Jiang_SciRep2016Raman, Spustaka_APLMater2025Contrast, Naylor_2DMat2017Large}. The flakes were subsequently transferred onto pre-patterned Si/SiO$_2$ substrates. The pre-patterned electrodes were fabricated by electron-beam lithography followed by deposition of 3--5~nm Pt. To minimize sample degradation, the heterostructures were assembled immediately after exfoliation using a dry-transfer technique. The resulting device architecture, Si/SiO$_2$/pre-patterned electrodes/WTe$_2$/h-BN/FLG, is schematically illustrated in Fig.~\ref{MFE_Fig1}(a).

Figure~\ref{MFE_Fig1}(b) illustrates the measurement geometry. The longitudinal resistance was measured along the crystallographic $a$ axis, parallel to the flake length and the direction of interlayer sliding~\cite{Kim_NatCommun2024Intrinsic}. The gate voltage was applied along the crystallographic $c$ axis to switch the polarization. The crystal structure is adapted from Ref.~\cite{Yang_JPCL2018Origin}. Four devices, D$_1$--D$_4$, were investigated for reproducibility and comparative analysis. Their thicknesses were tentatively estimated to be below 10~nm based on the optical contrast of WTe$_2$ flakes on 300-nm SiO$_2$/Si substrates~\cite{Na_Nanoscale2016tuning, Jiang_SciRep2016Raman, Spustaka_APLMater2025Contrast, Naylor_2DMat2017Large}.

Electrical transport measurements were performed using a low-frequency lock-in technique at 17.7~Hz, with an in-plane AC excitation current applied through the WTe$_2$ channel. The temperature dependence of the sheet resistance of the four devices is shown in Fig.~\ref{MFE_Fig1}(c), demonstrating metallic transport. The inset shows a representative optical micrograph of device D$_1$, with the individual layers indicated. Transport-detected ferroelectric hysteresis was measured by sweeping the top-gate voltage ($V_g$) while recording the in-plane sheet resistance ($R$) during forward (FS) and backward (BS) sweeps. Throughout this work, we use the term FE hysteresis referring to the hysteresis in sheet resistance rather than a conventional polarization--electric-field ($P$--$E$) loop. Figure~\ref{MFE_Fig1}(d) shows the gate-voltage-dependent resistance hysteresis for all four devices at $T=2$~K. This transport signature is consistent with FE switching reported in few-layer WTe$_2$~\cite{Fei_Nature2018Ferroelectric, Chiang_NanoLett2025Spatially, Yang_JPCL2018Origin, Sharma_SciAdv2019Room, Bai_NatCommun2025Sub, Liu_Nanoscale2019Vertical}.

To investigate the magnetic-field dependence and anisotropy of the FE hysteresis, magnetic fields were applied perpendicular to the sample plane ($B\parallel c$) and in-plane, parallel to the current ($B\parallel I$). The magnetic-field orientation was controlled using a rotatable sample stage integrated into a cryogenic superconducting magnet system. The excitation current was optimized to minimize Joule heating and measurement noise.

\section{Results}
\begin{figure*}
    \centering
    \includegraphics[width=0.6\linewidth]{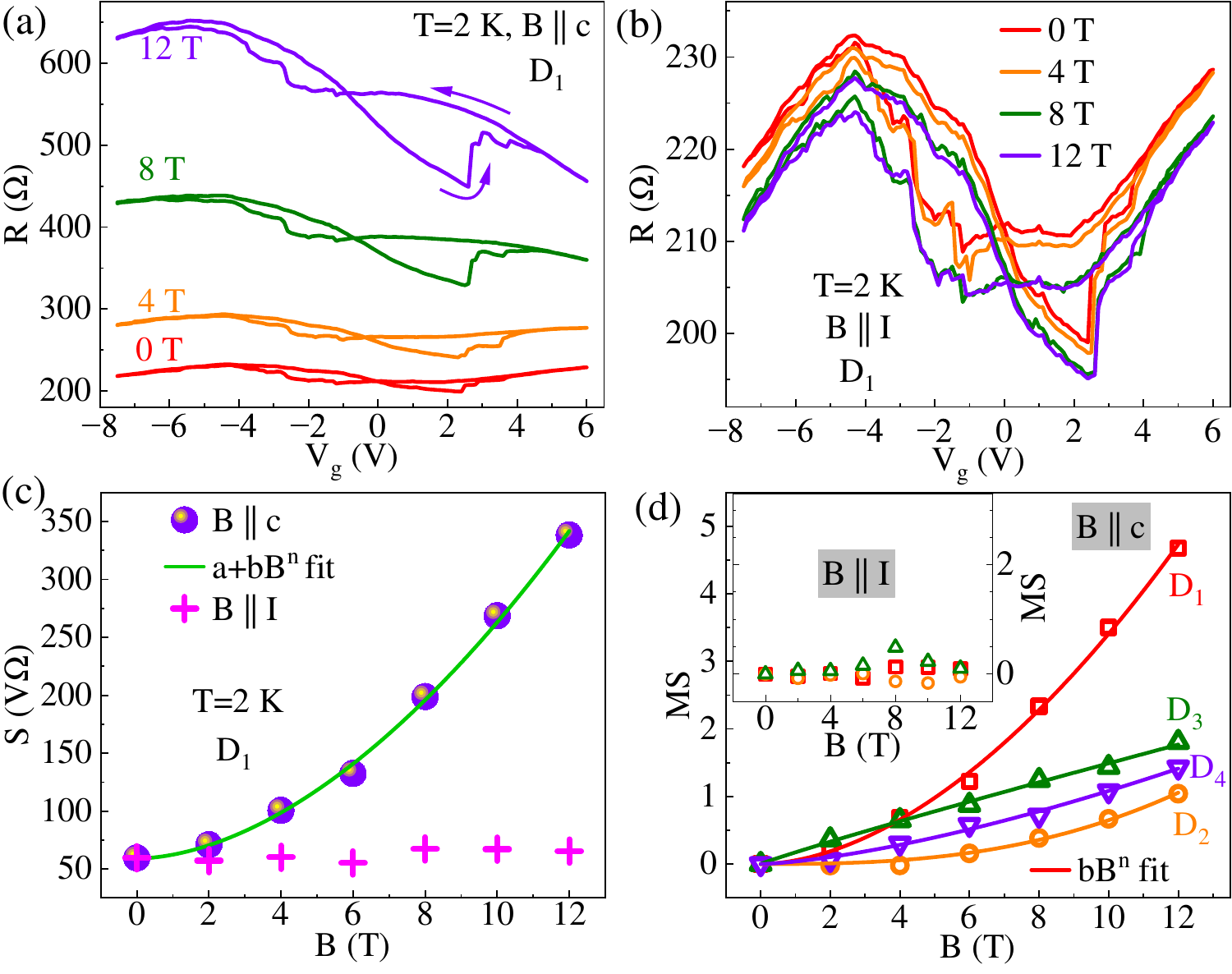}
\caption{Anisotropic FE hysteresis-area enhancement under magnetic field at $T=2$~K. FE hysteresis loops at different fixed (a) out-of-plane and (b) in-plane magnetic fields. (c) Hysteresis area $S$ as a function of magnetic field for both field orientations. Solid lines represent fits to $S(B)=a+bB^n$. (d) Normalized hysteresis-area change $MS$ as a function of out-of-plane magnetic field for all four devices (inset: in-plane response). Solid lines represent fits to $MS=b'B^n$.}
    \label{MFE_fig2}
\end{figure*}
{\it \textbf{Anisotropic FE hysteresis area enhancement under magnetic field:}}
The influence of magnetic-field strength and orientation on the FE hysteresis is summarized in Fig.~\ref{MFE_fig2}. Figures~\ref{MFE_fig2}(a) and (b) show the hysteresis loops under out-of-plane ($B\parallel c$) and in-plane ($B\parallel I$) magnetic fields, respectively for the device D$_1$. For the out-of-plane configuration, the resistance hysteresis measured under positive and negative magnetic fields were symmetrized by averaging the corresponding values, $[R(V_g,+B)+R(V_g,-B)]/2$, following the standard symmetrization procedure used in magnetotransport measurements ($V_g$ is the gate voltage). See Supplementary material~\footnote{Supplementary materials includes detailed discussions on (Section S1) Raw Hysteresis Loops Under Positive and Negative Magnetic Fields along out-of-plane c-axis for the device D$_1$; (Section S2) Magnetic field dependent FE hysteresis analysis of devices D$_2$–D$_4$ and corresponding branch resistance analysis; (Section S3)Ferroelectric hysteresis at different temperatures and at zero magnetic field for the device D$_1$}, Section: S1 for details. We note that, the hysteresis loops under positive and negative out-of-plane magnetic fields exhibits an even-in-field enhancement (the contrast between two resistive states corresponding to the two FE states). On the other hand, the enhancement in the hysteresis remains nearly unchanged under $B\parallel I$. Such contrasting response indicate a strong anisotropy in the magnetotransport associated with the two hysteretic resistance states.

To quantify the field-induced evolution of the hysteresis, we calculated the loop area by integrating the absolute resistance difference between the forward-sweep (FS) and backward-sweep (BS) branches. The resulting field dependence for device D$_1$ is shown in Fig.~\ref{MFE_fig2}(c). The out-of-plane field dependence is described by an empirical power-law relation, $S(B)=a+bB^n$, yielding $a=59.1\pm4.4$~V$\Omega$, $b=3.31\pm0.78$, and $n=1.79\pm0.09$. This power-law parameterization provides a compact description of the field dependence and enables comparison across devices. In contrast, the FE hysteresis area under $B\parallel I$ configuration remains unaffected (Fig.~\ref{MFE_fig2}(c)). The corresponding analysis for devices D$_2$--D$_4$ is presented in the Supplementary Material~\cite{Note1}, Section S2 (A). All four devices exhibit an enhancement of the hysteresis area under out-of-plane magnetic fields, with the field dependence described by similar power-law behaviour.

For comparison across devices, we define the normalized hysteresis-area change as $MS=[S(B)-S(0)]/S(0)$, analogous to the conventional magnetoresistance, $MR=[R(B)-R(0)]/R(0)$. The resulting $MS$ values are shown in Fig.~\ref{MFE_fig2}(d). All four devices exhibit an increase in $MS$ with out-of-plane magnetic field, described by $MS=b'B^n$. The extracted exponents are $n=1.79\pm0.07$, $2.67\pm0.20$, $0.94\pm0.04$, and $1.46\pm0.09$ for D$_1$--D$_4$, respectively. Despite the consistent enhancement, the exponent varies considerably among devices. In contrast, the in-plane response shown in the inset of Fig.~\ref{MFE_fig2}(d) is negligible for devices D$_1$--D$_3$, for which measurements were performed; the in-plane response of D$_4$ was not measured.



\begin{figure*}
    \centering
    \includegraphics[width=0.55\linewidth]{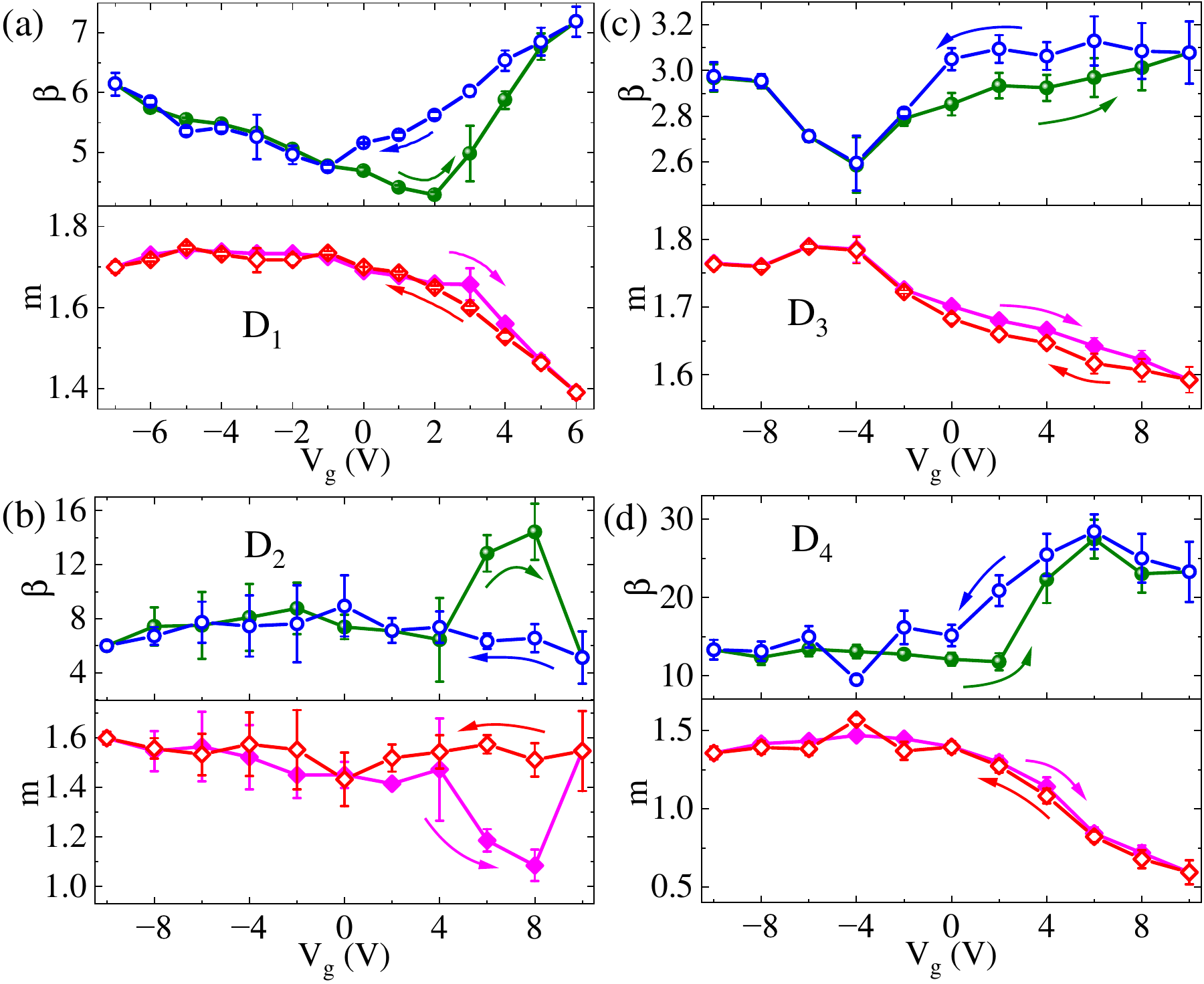}
 \caption{Gate-voltage dependence of the power-law fitting parameters obtained from branch-resolved magnetotransport measurements under out-of-plane magnetic field at $T=2$~K. The resistance change is fitted using $\Delta R(B,V_g)=\beta(V_g)B^{m(V_g)}$. Panels (a)--(d) correspond to devices D$_1$--D$_4$, respectively. Arrows indicate the gate-voltage sweep directions. Error bars represent uncertainties in the fitting parameters.}
    \label{MFE_fig3}
\end{figure*}

{\it \textbf{Branch-resolved magnetotransport:}}
The even-in-field enhancement and anisotropy of the FE hysteresis motivate us to examine the magnetic-field-dependent resistance of the individual hysteresis branches. Such branch-resolved analysis can reveal whether the enhancement arises from distinct magnetotransport responses of the two resistance states~\cite{Zhao_PRB2015anisotropic, thoutam2015temperature, ali2014large}. We therefore measured the resistance as a function of out-of-plane magnetic field at selected gate voltages, separately for the forward-sweep (FS) and backward-sweep (BS) branches. To isolate the field-induced resistance change, we define
$\Delta R(B,V_g)=R(B,V_g)-R(0,V_g)$,
where $R(0,V_g)$ is the corresponding zero-field resistance at the same gate voltage and on the same hysteresis branch. The resulting $\Delta R(B,V_g)$ was independently fitted for each branch using
$\Delta R(B,V_g)=\beta(V_g)B^{m(V_g)}$.
Representative resistance data and fits for all four devices are provided in the Supplementary Material~\cite{Note1}, Section S2 (B). The extracted parameters $\beta$ and $m$ are summarized in Fig.~\ref{MFE_fig3}.

The coefficient $\beta$ exhibits pronounced gate-voltage dependence and hysteretic behavior in all devices, particularly within the gate-voltage range of the zero-field FE hysteresis. This indicates that the two hysteresis branches exhibit distinct field-induced resistance changes. In contrast, the exponent $m$ shows comparatively weaker and less consistent hysteresis, suggesting that the amplitude of the magnetic-field-induced resistance change is more sensitive to the hysteresis branch than its field scaling. Therefore, at a given $V_g$, the higher-resistance branch exhibits a larger field-induced resistance increment. Consequently, the resistance difference between the two branches increases with magnetic field, providing a direct explanation for the enhancement of the hysteresis area.

{\it \textbf{Evolution of magnetic-field-enhanced FE hysteresis with temperature:}}
\begin{figure}
    \centering
    \includegraphics[width=0.95\linewidth]{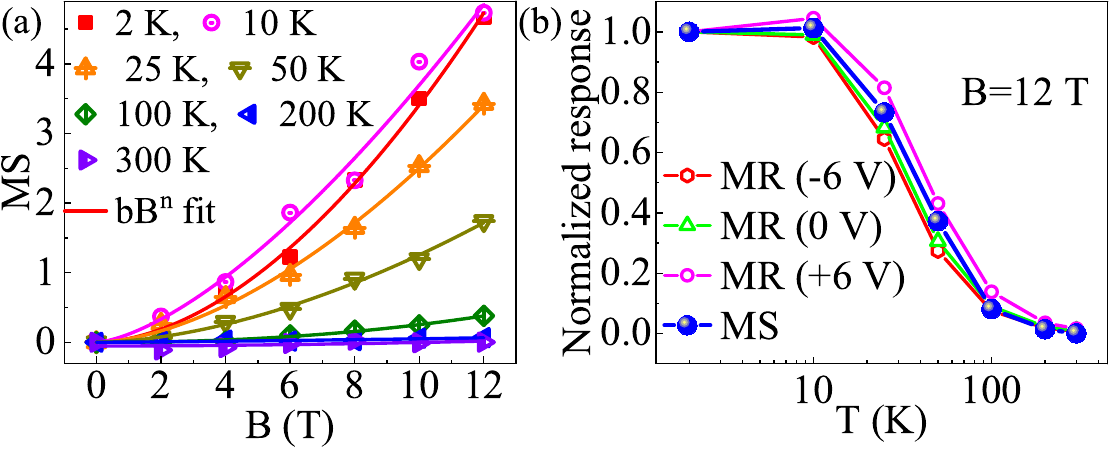}
    \caption{Temperature dependence of the magnetic-field-enhanced FE hysteresis in device D$_1$. (a) Normalized hysteresis-area change ($MS$) as a function of magnetic field at different temperatures. Solid lines represent fits to $MS=b'B^n$. (b) Temperature dependence of $MS$ and magnetoresistance (MR) at three representative gate voltages ($V_g=\pm6$~V and 0~V), measured at 12~T and normalized to their respective values at $T=2$~K.}
    \label{MFE_Fig4}
\end{figure}
Having established the distinct magnetotransport responses of the two hysteresis branches, we next examine the temperature dependence of the magnetic-field-induced hysteresis enhancement in device D$_1$, as shown in Fig.~\ref{MFE_Fig4}. Figure~\ref{MFE_Fig4}(a) presents the normalized hysteresis-area change ($MS$) as a function of magnetic field at different temperatures. The data are fitted using $MS=b'B^n$, as shown by the solid lines. The enhancement gradually decreases with increasing temperature and becomes negligible near room temperature, following a trend similar to the temperature dependence of magnetoresistance in WTe$_2$~\cite{Manna_PRB2026Highorder, Wang_PRB2015_Origin, Sharma_prb17}.

To examine this correspondence, Fig.~\ref{MFE_Fig4}(b) compares the temperature dependence of $MS$ and $MR$ at three representative gate voltages ($V_g=\pm6$~V and 0~V). Both quantities are evaluated at 12~T and normalized to their respective values at 2~K. The similar temperature evolution of $MS$ and $MR$ suggests that the magnetic-field-induced enhancement of the transport-detected FE hysteresis is closely associated with the temperature-dependent magnetotransport response.

\section{Discussion}

The combined magnetic-field, branch-resolved, and temperature-dependent measurements reveal a close association between the magnetic-field-induced enhancement of transport-detected ferroelectric (FE) hysteresis and the distinct magnetotransport responses of the two hysteresis branches. The higher-resistance branch exhibits a larger field-induced resistance increment, increasing the resistance contrast between the branches while largely preserving the overall hysteresis-loop shape. This enhancement is strongly anisotropic, occurring predominantly under out-of-plane magnetic fields, and is reproducible across multiple devices. Furthermore, its temperature evolution closely follows that of the magnetoresistance (MR), suggesting that the underlying electronic transport response plays an important role in the observed hysteresis enhancement.

A plausible physical picture can be developed from the interplay between ferroelectricity and the electronic structure of few-layer WTe$_2$. Ferroelectric switching in bilayer WTe$_2$ originates from interlayer sliding, which produces two polarization states with distinct electronic configurations~\cite{delaBarrera_NatCommun2021Direct, Fei_Nature2018Ferroelectric,Yang_JPCL2018Origin,Tang_PRL2023Sliding}. It has already been demonstrated that the spontaneous polarization depends strongly on carrier density, with layer-polarized low-energy electronic states contributing directly to the polarization~\cite{delaBarrera_NatCommun2021Direct}. Meanwhile, the large positive MR of semimetallic WTe$_2$ is commonly associated with near compensation between electron and hole populations, making its magnetotransport sensitive to changes in carrier balance~\cite{Y.Wang,Lv_EPL2015Perfect,M.N.Ali2}. Accordingly, the two FE states may exhibit different carrier compensation and, consequently, distinct magnetotransport responses. Such differences could account for the unequal field-induced resistance increments and the resulting enhancement of the hysteresis area. This interpretation does not require a direct magnetic-field-induced modification of the FE polarization states themselves.

Another notable feature is that the hysteresis enhancement is predominantly even with respect to magnetic-field reversal, as demonstrated by the nearly identical responses under positive and negative magnetic fields (Supplementary Material~\cite{Note1}, Section S1). This is expected in cases with where the time reversal symmetry is not broken but the inversion symmetry is broken. However, it motivates consideration of higher-order magnetoelectric coupling. Within a phenomenological free-energy description, higher-order coupling between ferroelectric polarization and magnetic field may contribute to the observed even-in-field response~\cite{Spaldin_NatMater2019Advances,Anderson_PRL1965Symmetry,Benedek_JMaterChemC2016Ferroelectric}. Although the present transport measurements do not directly establish the microscopic magnetoelectric coupling, the observed field symmetry is compatible with such higher-order contributions. This may provide a complementary phenomenological framework for interpreting the magnetic-field dependence of transport-detected FE hysteresis in few-layer WTe$_2$.

A related manifestation of magnetoelectric coupling has recently been reported in few-layer WTe$_2$~\cite{Ye_PRB2025Current}. In that study, a Berry-connection-polarizability (BCP)-driven nonlinear magnetoelectric response was associated with current-induced out-of-plane magnetization and an anomalous Hall effect, with pronounced anisotropy arising from the intrinsic electronic structure and BCP tensor of WTe$_2$. Although this nonlinear response is distinct from the hysteresis enhancement investigated here, both phenomena highlight the role of WTe$_2$'s anisotropic electronic structure in coupling electronic transport to magnetic and electric responses. Our results provide a transport-based perspective on this interplay through the distinct magnetotransport responses of the two FE polarization branches.

\section{Conclusion}

We establish a close connection between anisotropic magnetotransport and the magnetic-field-induced enhancement of transport-detected ferroelectric hysteresis in few-layer WTe$_2$. The phenomenon is reproducible over several few layer devices. The distinct field-induced resistance changes of the two ferroelectric hysteresis branches, together with the similar temperature evolution of the hysteresis enhancement and magnetoresistance, point to the underlying electronic transport response as a key factor governing the observed effect. Although the microscopic origin of the nonlinear magnetoelectric coupling mechanism remains to be established, these findings highlight the potential of anisotropic magnetotransport to modulate the electrical contrast between ferroelectric states, providing a perspective on the interplay between ferroic order and electronic transport in two-dimensional quantum materials.

\section{Acknowledgments}

The authors acknowledge IIT Kanpur, the Department of Science and Technology (DST), India, [Order No. DST/NM/TUE/QM-06/2019 (G)] and DST-FIST (DST/PHY/2022 523) for financial support. 

\bibliography{cite}

\end{document}